%% file: paper.tex
\documentclass[twocolumn,showpacs,aps,prd,floatfix]{revtex4}

\usepackage{graphicx}
\usepackage{dcolumn}
\usepackage{amsmath}
\usepackage{epsfig}
\usepackage{bm}

\input pubboard/babarsym

\newcommand{\kevcc}{\ensuremath{{\mathrm{\,ke\kern -0.1em V\!/}c^2}}\xspace}

\def\pcm {\ensuremath{p^*}\xspace}
\def\Lc {\ensuremath{\Lambda_c^+}\xspace}

\def\Lz {\ensuremath{\Lambda}\xspace}

\def\Sz {\ensuremath{\Sigma^0}\xspace}
\def\LKKz {\ensuremath{\Lz \Kzb \Kp }\xspace}
\def\LKKs {\ensuremath{\Lz \KS \Kp }\xspace}
\def\SKKs {\ensuremath{\Sz \KS \Kp }\xspace}

\def\LcLKKs {\ensuremath{\Lc\to\LKKs}\xspace}
\def\LcLKKz {\ensuremath{\Lc\to\LKKz}\xspace}
\def\LcSKKs {\ensuremath{\Lc\to\SKKs}\xspace}
\def\pKpi {\ensuremath{\proton \Km \pip}\xspace}
\def\LcpKpi  {\ensuremath{\Lc\to\pKpi}\xspace}
\def\pKs {\ensuremath{\proton \KS}\xspace}
\def\LcpKs  {\ensuremath{\Lc\to\pKs}\xspace}
\def\Qval {\ensuremath{Q}-value\xspace}

\newcommand{\BABARPubYear}    {05}
\newcommand{\BABARPubNumber}  {032}

\newcommand{\SLACPubNumber} {11331}

\def\figurebox#1#2#3{%
    \def\arg{#3}%
    \ifx\arg\empty
    {\hfill\vbox{\hsize#2\hrule\hbox to #2{\vrule\hfill\vbox to #1{\hsize#2\vfill}\vrule}\hrule}\hfill}%
    \else
    {\hfill\epsfbox{#3}\hfill}%
    \fi}

\begin{document}


\begin{flushleft}
\babar-PUB-\BABARPubYear/\BABARPubNumber\\
SLAC-PUB-\SLACPubNumber\\
\end{flushleft}

\title{
{\large \bf
A Precision Measurement of the $\Lc$ Baryon Mass} 
}

\input pubboard/authors_jun2005
\input PRDconst.tex

\date{\today}

\begin{abstract}
The \Lc baryon mass is measured using \LcLKKs and \LcSKKs decays
reconstructed in 232\invfb of data collected with the \babar\
detector at the PEP-II asymmetric-energy \epem storage ring. The \Lc mass is
measured to be $\constCombMass\mevcc$.  The dominant systematic
uncertainties arise from the amount of material in
the tracking volume and from the magnetic field strength.
\end{abstract}


\maketitle

\section{Introduction}

\label{sec:introduction}

\input introduction.tex

\section{Analysis Method}

\label{sec:method}

\input method.tex

\section{The \babar\ Detector}

\label{sec:detector}

\input detector.tex

\section{Data Sample}

\label{sec:samples}

\input samples.tex

\section{Event Selection}

\label{sec:selection}

\input selection.tex

\section{Signal Fits}

\label{sec:signal}

\input signal.tex

\section{Cross-checks}

\label{sec:crosschecks}

\input crosschecks.tex

\section{Systematic Uncertainties}

\label{sec:systematics}

\input systematics.tex

\section{Summary}

\label{sec:summary}

\input summary.tex


\begin{acknowledgments}
\input pubboard/acknowledgements.tex
\end{acknowledgments}

\end{document}

%% file: pubboard/authors_jun2005.tex
%
\author{B.~Aubert}
\author{R.~Barate}
\author{D.~Boutigny}
\author{F.~Couderc}
\author{Y.~Karyotakis}
\author{J.~P.~Lees}
\author{V.~Poireau}
\author{V.~Tisserand}
\author{A.~Zghiche}
\affiliation{Laboratoire de Physique des Particules, F-74941 Annecy-le-Vieux, France }
\author{E.~Grauges}
\affiliation{IFAE, Universitat Autonoma de Barcelona, E-08193 Bellaterra, Barcelona, Spain }
% ...
\collaboration{The \babar\ Collaboration}
\noaffiliation

%% file: PRDconst.tex
\def\constCombCorr{0.355} 
\def\constCombMass{2286.46\pm0.14} 
\def\constDistortSagitta{11} 
\def\constLKKsDistortTwo{55} 
\def\constLKKsFrac{(83 x  3)\%} 
\def\constLKKsHWHM{  2.55 x  0.06} 
\def\constLKKsMassOffset{61} 
\def\constLKKsMean{2286.44 x  0.04} 
\def\constLKKsNarrow{  2.08 x  0.07} 
\def\constLKKsPRDAlign{ 23} 
\def\constLKKsPRDFitProc{ 38} 
\def\constLKKsPRDMagnet{ 68} 
\def\constLKKsPRDMassChg{ 27} 
\def\constLKKsPRDMaterial{ 83} 
\def\constLKKsPRDSolenoid{ 60} 
\def\constLKKsPRDTotal{144} 
\def\constLKKsResult{2286.501\pm0.042(stat.)\pm0.144(syst.)\mevcc} 
\def\constLKKsWide{  6.39 x  1.22 } 
\def\constLKKsYield{4627 x   84} 
\def\constMassResultks{  497.56\pm0.04(stat.)\pm0.26(syst.)} 
\def\constMassResultlz{1115.68\pm0.01(stat.)\pm0.04(syst.)} 
\def\constSKKsDistortTwo{30} 
\def\constSKKsFrac{$(100\pm 0)\%$} 
\def\constSKKsHWHM{  2.41 x  0.22} 
\def\constSKKsMassOffset{18} 
\def\constSKKsMean{2286.29 x  0.18} 
\def\constSKKsNarrow{  2.04 x  0.18} 
\def\constSKKsPRDAlign{ 13} 
\def\constSKKsPRDEScale{ 46} 
\def\constSKKsPRDFitProc{ 71} 
\def\constSKKsPRDMagnet{ 29} 
\def\constSKKsPRDMassChg{ 58} 
\def\constSKKsPRDMaterial{ 50} 
\def\constSKKsPRDSolenoid{ 30} 
\def\constSKKsPRDTotal{126} 
\def\constSKKsResult{2286.303\pm0.181(stat.)\pm0.126(syst.)\mevcc} 
\def\constSKKsYield{ 264 x  20} 
\def\constpKpiCorrection{211} 
\def\constpKpiResult{ 2286.39\pm0.02(stat.)\pm0.45(syst.)\mevcc} 
\def\constpKsCorrection{145} 
\def\constpKsResult{ 2286.36\pm0.03(stat.)\pm0.43(syst.)\mevcc} 

%% file: introduction.tex
The invariant masses of the stable charmed hadrons are currently
reported by the Particle Data Group (PDG) with a precision of about
0.5--0.6\mevcc~\cite{PDG}. The best individual measurements have a
statistical and systematic precision of about $0.5$\mevcc and use data
samples of a few hundred events.  The
\babar~data contains large samples of many charmed-hadron
decays and, due to the excellent momentum and vertex resolution in
\babar, many of the decay modes can be reconstructed with an
event-by-event mass uncertainty of a few \mevcc. We can therefore
significantly improve the precision on the charm-hadron mass
measurements.

In this analysis we present a precision measurement of the \Lc
mass. The measurement is based on the reconstruction of the decay
modes \LcLKKs and \LcSKKs \cite{conjugates}. Because almost all of the
\Lc invariant mass in these decays results from the well-known rest-mass 
values of the \Lc decay products, the systematic uncertainty in
the reconstructed mass is significantly reduced compared to the
precision obtained in other decay modes. Large samples of
\LcpKpi and \LcpKs decays are used for cross-checks and studies of
systematic uncertainties.

%% file: method.tex
The measurement of invariant mass relies on precise and unbiased
measurements of particle three-momenta. If the chosen decay mode
contains any photons, unbiased energy and position measurements in the
electromagnetic calorimeter (EMC) are also necessary. The momentum
measurement depends on a well-aligned detector, precise knowledge of
the magnetic field and material distribution in the tracking volume,
and mass-dependent corrections for the energy loss of charged
particles passing through the detector material. All three
requirements are met to a large degree by the \babar\ detector and its
event reconstruction algorithms. We quantify residual systematic
effects in studies of various control samples.  To minimize such
contributions to the systematic uncertainty, we choose to measure the
\Lc mass by using decay modes with a low \Qval, where the \Qval for a
decay $a\to b+c+\ldots$ is defined as
\begin{equation*}
Q=m(a)-m(b)-m(c)-\ldots
\end{equation*}
Uncertainties related to track reconstruction, such as those involving
energy-loss correction or magnetic field strength, tend to scale with the
\Qval.

The main signal mode used in this analysis is \LcLKKs, which has a
\Qval of 177.9\mevcc calculated using the PDG \Lc~mass. The branching 
ratio $\Gamma(\LcLKKz)/\Gamma(\LcpKpi)$ was measured by CLEO to be
$0.12\pm 0.02 \pm 0.02$~\cite{Ammar:1995je}.  BELLE observed that
about $25\%$ of these decays proceed through
$\Xi(1690)^0\Kp$~\cite{Abe:2001mb}. Since the $\Xi(1690)^0$ baryon has
a width of several \mevcc and its mass is not well-known, it does not
help in constraining the \Lc mass. We reconstruct the \Lz and \KS only
in their charged decay modes, $\Lz\to\proton\pim$ and $\KS\to\pipi$,
which account for about 44\% of \LcLKKs decays. To obtain the most
precise \Lc mass value, the long-lived particles are reconstructed
with their mass values constrained to their respective PDG values.

A second low \Qval \Lc decay mode, \LcSKKs, has also been studied. With the
\Qval of 100.9\mevcc it has the potential for an even lower systematic
uncertainty. The results of this study indicates that the branching
fraction for this decay mode is significantly smaller than for the
\LcLKKs decay mode, which makes the statistical precision worse than
the total precision of the \LcLKKs mode. The decay also requires the
measurement of a photon from the $\Sz\to\Lz\gamma$ decay and this
introduces an additional systematic uncertainty.

For \LcLKKs decays, the event-by-event uncertainty on the invariant
mass is about 2\mevcc; hence, with a few thousand reconstructed signal
events, the statistical uncertainty is below 50\kevcc and the dominant
uncertainty is systematic. The major part of the systematic
uncertainty is estimated directly from the data by redoing the track
fits with different assumptions on the amounts of detector material or
the magnetic field strength, and measuring how much the invariant mass
value changes with each assumption. The candidate selection criteria
have been optimized on simulated events to minimize the expected
systematic uncertainty.

To check that the procedure for estimating the systematic uncertainty
is reasonable, we study large samples of \Lz and \KS decays.  The goal
is to ensure that the measured mass values are consistent with the PDG
values within the systematic uncertainty we estimate using the
same procedure as for the signal mode.  Large samples of
\LcpKpi and \LcpKs decays are also used and give invariant mass
measurements that are consistent with the signal when their larger
systematic uncertainties are taken into account.

%% file: detector.tex
The \babar\ detector is described in detail
elsewhere~\cite{ref:babar}.  The momenta of charged particles are
measured with a combination of a five-layer silicon vertex tracker
(SVT) and a 40-layer drift chamber (DCH) in a 1.5-T solenoidal
magnetic field.  The momentum resolution is measured to be
$\sigma(\pt)/\pt=0.0013(\pt/\gevc)\oplus 0.0045$. A detector of
internally reflected Cherenkov radiation (DIRC) is used in charged
particle identification.  Kaons and protons are identified with
likelihood ratios calculated from $dE/dx$ measurements in the SVT
and DCH, and from the observed pattern of Cherenkov light in the DIRC.
A finely segmented CsI(Tl) electromagnetic calorimeter (EMC) is used
to detect and measure photons and neutral hadrons, and to identify
electrons.  The instrumented flux return (IFR) contains resistive
plate chambers for muon and long-lived neutral-hadron identification.
For event simulation we use the Monte Carlo generator EVTGEN \cite{evtgen}
with a full detector simulation that uses GEANT4 \cite{geant4}.

The most critical component of this analysis is the quality of the
track reconstruction. In order to maximize tracking
efficiency, the track-finding algorithm is based
on tracks found by the trigger system and by stand-alone track
reconstruction in the SVT and in the DCH. Once a track has been found,
the track parameters are determined using a Kalman filter
algorithm~\cite{chep2000}, which makes optimal use of the hit
information and corrects for energy loss and multiple scattering in
the material traversed and for inhomogeneities in the magnetic
field. The material-traversal corrections change the track momentum
according to the expected average energy loss and increase the
covariance for track parameters to account for both multiple
scattering and the variance in the energy loss. The energy loss
depends on the particle velocity, therefore each track fit is
performed separately for five particle hypotheses: electron, muon,
pion, kaon, and proton.  A simplified model of the \babar\ detector
material distribution is used in the Kalman filter algorithm in order
to maintain reasonable execution time. The main layers of material
traversed by a particle originating from the interaction point are the
beam pipe at a radius of 2.5\,cm, consisting of about 1.4-mm of beryllium
and 1.5-mm of cooling water; five layers of 300-$\mu$m-thick silicon
detectors at radii from 3.3\,cm to 15\,cm; a 2-mm-thick carbon-fiber
tube at a radius of 22\,cm supporting the SVT and beam-line magnets;
and the inner wall of the DCH at a radius of 24\,cm, which is a 1-mm
thick beryllium tube.  Part of the support structure for the silicon
detectors is modeled by increasing the thickness of each layer in the
silicon detector by 60\mum, while the effect of the DCH gas is modeled
as a series of discrete material contributions.  Detailed knowledge of
the magnetic field is also essential to the track reconstruction. This
is discussed in Section~\ref{sec:magfield} below.

%% file: samples.tex
The data sample used for the \Lc mass measurement comprises an
integrated luminosity of 232\invfb collected from \epem collisions at
or 40 \mev below the \FourS resonance. For the studies of \Lz and \KS
decays only a small subsample of the data is used due to the high
production rates of these hadrons.  Studies of simulated events are
performed using Monte Carlo samples of generic $\epem\to\BB$ and
$\epem\to q\bar{q} (q=u,d,s,c)$ continuum events with an integrated
luminosity equivalent to 240 to 275\invfb. More than 230000 simulated
\LcLKKs decays and 60000 simulated \LcSKKs decays are used for studies of
systematic uncertainty.

%% file: selection.tex
\subsection{$\protect\bm{\LcLKKs}$ Selection}

\label{sec:lkksSelection}

The \LcLKKs signal is reconstructed using only the charged two-body
decay modes of the \Lz and \KS hadrons. We form \Lz candidates from
two tracks, one of which must be identified as a proton, and, after
fitting to a common vertex, we require the combined invariant mass to be
between 1106 and 1125\mevcc. For surviving candidates, the two tracks
are fit to a common vertex with the invariant mass constrained to the
PDG \Lz mass. The probability of this
mass-constrained vertex fit is required to be above
$10^{-3}$. Similarly, a \KS candidate is formed from two tracks,
neither of which belongs to the \Lz candidate, with a combined
invariant mass between 460 and 530\mevcc.  For surviving
\KS candidates, the two daughter tracks are fit to a common vertex
with the mass constrained to the PDG
\KS mass and the fit probability is required to be higher than
$10^{-3}$. The \Lz and \KS candidates are then combined with a fifth
track, identified as a charged kaon, in a fit to a common vertex to
form a \Lc candidate. The \Lc candidate must have an invariant mass
between 2250\mevcc and 2330\mevcc. The probability of the vertex fit
should be greater than $10^{-3}$. To suppress combinatorial
background, the signed decay length of a \KS candidate is required to
be larger than three times its estimated uncertainty.  The signed
decay length is defined as the distance between the \Lc and \KS
candidate along the \KS momentum in the CM frame.  To further suppress
background, which results mainly from \B decays, the momentum (\pcm)
of the \Lc candidate in the \epem center-of-mass frame (CM) is
required to be at least 2\gevc. This requirement also helps to reduce
systematic uncertainties that affect mainly low-momentum tracks. The
selection efficiency, not including branching fractions, is about 15\%
for \LcLKKs decays with
\Lc CM momentum larger than 2\gevc. The background is suppressed
sufficiently to not be an issue for the \Lc mass measurement.

\subsection{$\protect\bm{\LcSKKs}$ Selection}

\label{sec:skksSelection}

The \LcSKKs mode is reconstructed from $\Sz\to\Lz\gamma$, and \Lz and
\KS hadrons decaying into two charged particles. The \Lz and \KS
candidates are formed in the same way as in the \LcLKKs selection.  A
\Sz candidate is formed by combining a \Lz candidate with a photon and
requiring the combined invariant mass to be between 1184 and
1196\mevcc. A photon candidate is defined as an energy deposit in the
EMC of at least 30\mev that is not associated with any track and has a
lateral moment \cite{LAT} of its shower energy deposition of less than
0.8. If several photons can be combined with a \Lz to form \Sz
candidates, only the candidate with the most energetic photon is
retained in order to avoid double counting. The \Sz candidates are fit
with their mass constrained to the PDG mass and are combined with \KS
and
\Kp candidates to form \Lc candidates that must satisfy
the same invariant mass, vertex probability, and \pcm requirements as
\LcLKKs candidates. The selection efficiency, not including branching
fractions, is estimated to be about 8\% for
\LcSKKs decays with \Lc CM momentum larger than 2\gevc.

From simulation we expect that 18\% of the \LcSKKs decays are
reconstructed with the wrong photon in the \Sz candidate. Due to the
low energy of the photon and the mass constraint on the \Sz candidate,
these \Lc candidates still have the correct mass on average, but
the mass resolution of these candidates is significantly degraded.

\subsection{$\protect\bm{\LcpKpi}$ Selection}

\label{sec:pKpiselect}

\LcpKpi candidates are formed from three tracks identified as
a proton, a kaon, and a pion with a combined invariant mass between
2240 and 2330\mevcc. The tracks are fit to a common vertex and the
probability of the vertex fit is required to be greater than $10^{-3}$.
The signal-selection efficiency is about 42\% and depends on the
\Lc momentum.

\subsection{$\protect\bm{\LcpKs}$ Selection}

\label{sec:pKsselect}

For the \LcpKs decay mode, the \KS candidates are required to satisfy
the same criteria as in the \LcLKKs mode, but with the further
constraint that the decay angle $\theta$ of the \pip in the \KS rest
frame with respect to the \KS line-of-flight must satisfy
$|\cos\theta|<0.97$. This removes contamination from $\gamma$
conversions. The \KS candidates are combined with tracks identified as
protons in a fit to a common vertex and the resulting
\Lc candidates are required to have an invariant mass between 2240 and
2330\mevcc. The probability of the vertex fit is required to be above
$10^{-3}$.  The signal selection efficiency is about 41\% and depends
on the \Lc momentum.

\subsection{$\protect\bm{\Lz\to\proton\pim}$ and $\protect\bm{\KS\to\pipi}$ Selection}

\label{sec:lzksselect}

For the cross check studies in Section~\ref{sec:crosschecks}, large
samples of \Lz and \KS decays are reconstructed using similar
criteria as those used to select \Lz and \KS candidates for
\LcLKKs decays.  The mass-constrained vertex fits are replaced with
geometric vertex fits with the requirement of a fit probability greater
than $10^{-2}$. For both \Lz and \KS candidates the signed decay length
calculated with respect to the \epem interaction point is required to
be larger than three times its uncertainty.

%% file: signal.tex
\begin{figure}
\centerline{\includegraphics{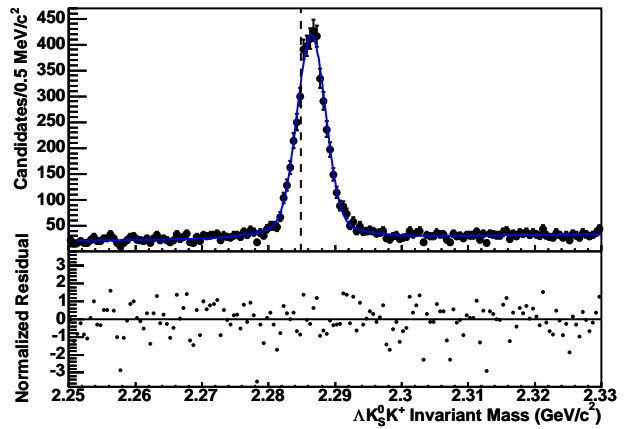}}
\caption{\label{fig:lkksSignal} Invariant mass distribution 
for \LcLKKs candidates. The lower part of the figure shows the
 normalized fit residuals. The dashed line indicates the present PDG
 value.}
\end{figure}

The invariant mass distribution for the \LcLKKs  candidates is
shown in Fig.~\ref{fig:lkksSignal}.  A clear \Lc signal peak is
observed.  A binned maximum likelihood fit to the mass distribution
is performed using a sum of two Gaussians with a common mean for the
\Lc signal function. The background is described by a linear function
as suggested by simulation studies. The fit parameter values are
given in Table~\ref{tab:signal}. Note that the uncertainty on the mean
mass is statistical only and a correction for underestimated energy
loss described in the next section has not been applied to the fitted
mass.

\begin{table}
\caption{\label{tab:signal} Fit parameter values for the \LcLKKs and \LcSKKs 
signals together with the half-width at half-maximum (HWHM) calculated
from these values.  A correction to the mass values for underestimated
energy loss has not been applied.}
\begin{ruledtabular}
\begin{tabular}{lD{x}{\pm}{-1}D{x}{\pm}{-1}}
Parameter& \multicolumn{1}{c}{\LcLKKs} & \multicolumn{1}{c}{\LcSKKs} \\ \hline
Fitted mass (\mevcc)    & \constLKKsMean   & \constSKKsMean  \\
Signal yield (events)& \constLKKsYield  & \constSKKsYield \\
Narrow width (\mevcc) & \constLKKsNarrow & \constSKKsNarrow\\
Broad width (\mevcc)   & \constLKKsWide   & \multicolumn{1}{c}{~~---}  \\
Narrow fraction       & \constLKKsFrac   & \multicolumn{1}{c}{~~---}  \\
HWHM (\mevcc)         & \constLKKsHWHM   & \constSKKsHWHM 
\end{tabular}
\end{ruledtabular}
\end{table}

\begin{figure}
\centerline{\includegraphics{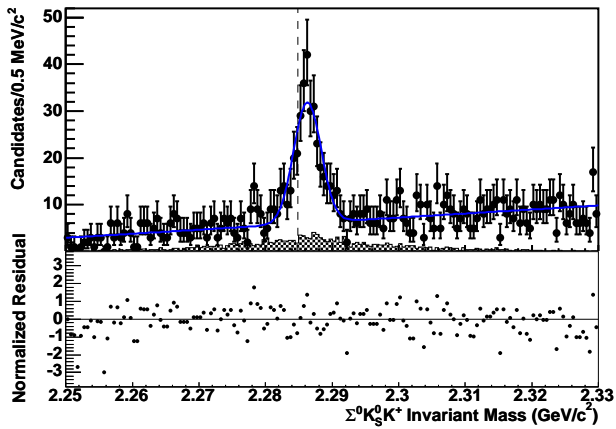}}
\caption{\label{fig:skksSignal} Invariant mass of reconstructed \LcSKKs 
candidates.  The hashed histogram is the expected contribution from signal
events with a wrongly assigned photon. The lower part of the figure
shows the normalized fit residuals. The dashed line indicates the present PDG value.
}
\end{figure}

The invariant mass distribution for the \LcSKKs candidates is
shown in Fig.~\ref{fig:skksSignal}.  A small but significant \Lc
signal peak is observed. The figure also shows the expected background
under the \Lc peak from \LcSKKs decays with a correct \Lz but a wrong
photon used in candidate selection. A binned maximum likelihood fit
of the mass distribution is performed using a single Gaussian for the
\Lc signal. The background is described by a linear function. The
wrongly-reconstructed \LcSKKs candidates are absorbed into the signal
and background because they peak at the \Lc mass. The fit parameter values
are given in Table~\ref{tab:signal}.

\begin{figure}
\centerline{\includegraphics{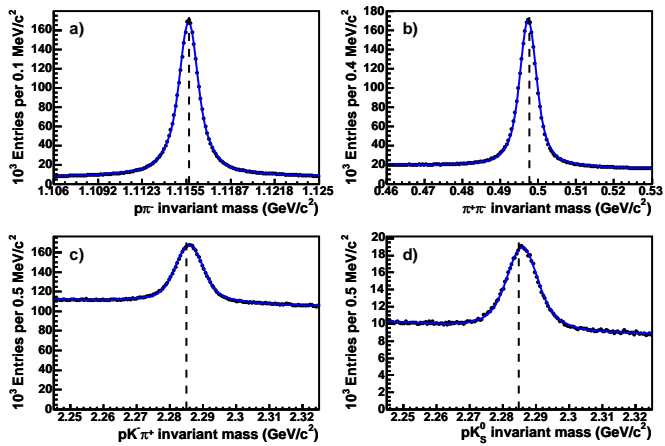}}
\caption{\label{fig:crossSignal} Invariant mass distribution of candidates
used in cross-checks of systematic uncertainty. a) $\Lz\to\proton\pim$ decays. b)
$\KS\to\pipi$ decays. c) \LcpKpi decays.  d) \LcpKs decays. The dashed
lines show the PDG mass values for these hadrons. A correction to
the mass distributions for underestimated energy loss has not been applied.}
\end{figure}

The invariant-mass distributions for the four control modes are shown
in Fig.~\ref{fig:crossSignal}. The \Lz and \KS signals are fit to a
sum of three Gaussians with common mean, while the two \Lc signals are
fit to a sum of two Gaussians with common mean. The background in
all four cases is modeled with a second-degree polynomial. The fit yields,
mass, and signal RMS values are listed in
Table~\ref{tab:crossSignal}. The fitted mass values for the \Lz and \KS
are significantly below the PDG values. This is mainly due
to an underestimation of the energy loss in the detector material and
is described in more detail in the next section.

\begin{table*}
\caption{\label{tab:crossSignal} Fit parameter values and calculated HWHM for the four 
decay modes used for cross-checks. A correction to each mass value for
underestimated energy loss has not been applied.}
\begin{ruledtabular}
\begin{tabular}{lD{x}{\pm}{-1}D{x}{\pm}{-1}D{x}{\pm}{-1}D{x}{\pm}{-1}}
Parameter& \multicolumn{1}{c}{$\Lz\to\proton\pim$} & \multicolumn{1}{c}{$\KS\to\pipi$} & 
\multicolumn{1}{c}{\LcpKpi} & \multicolumn{1}{c}{\LcpKs} \\
\hline PDG mass (\mevcc) &1115.683 x 0.006 &
497.648 x 0.022 & 2284.900 x  0.600 & 2284.900 x 0.600 \\
\input PRDtables/CrossSignals
\end{tabular}
\end{ruledtabular}
\end{table*}

%% file: PRDtables/CrossSignals.tex
Fitted mass (\mevcc) & 1115.660 x 0.001 & 497.305 x 0.002 & 2285.845 x 0.013 & 2285.876 x 0.023 \\
Signal yield (events) & 3192700 x 5800 & 2463900 x 4900 & 1449300 x 5300 & 243700 x 1900 \\
HWHM (\mevcc) & 0.853 x 0.002 & 2.715 x 0.005 & 5.147 x 0.014 & 5.613 x 0.046 \\

%% file: crosschecks.tex
The four control mode samples discussed above are used to understand
the accuracy to which particle masses can be measured in \babar.

\subsection{Material Dependence}

\begin{figure*}
\centerline{\includegraphics{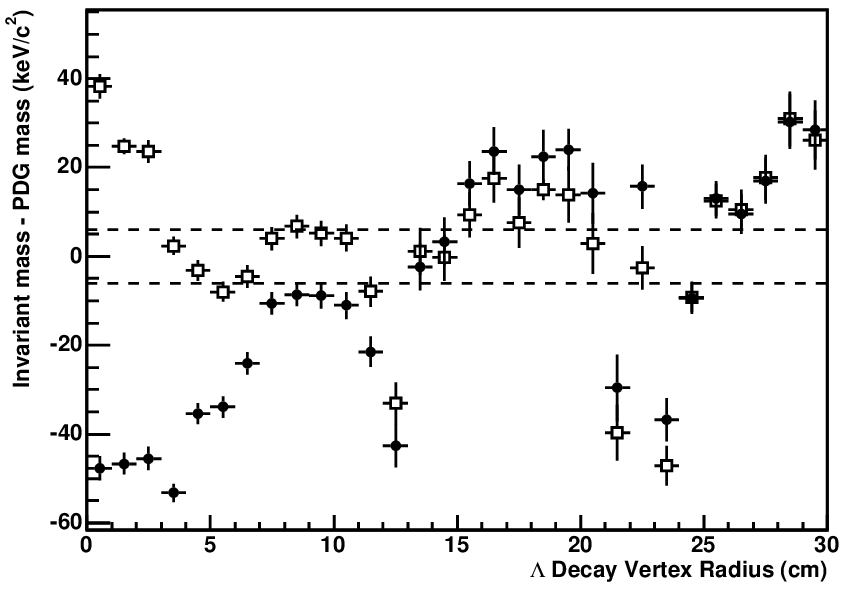} \includegraphics{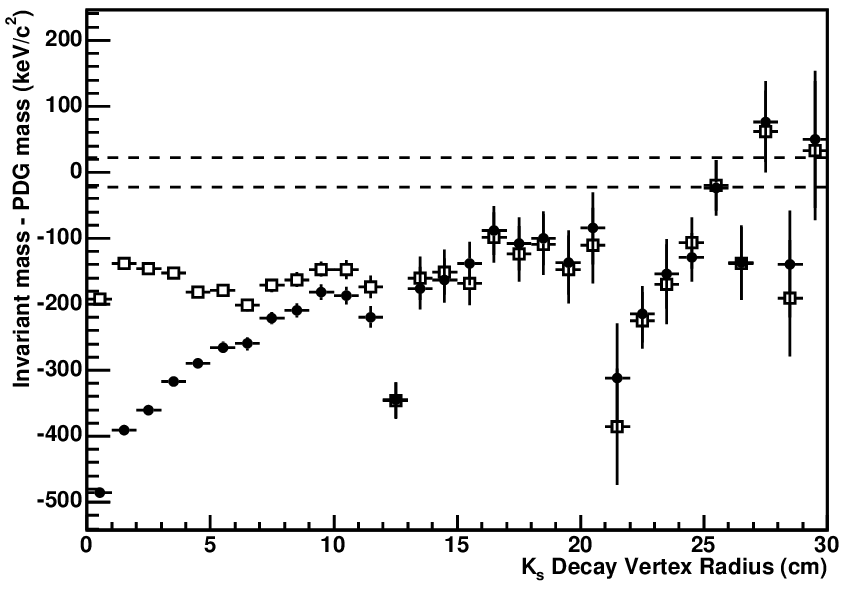}}
\caption{\label{fig:lzRadius} Fitted mass values for $\Lz\to\proton\pim$ 
(left) and $\KS\to\pipi$ (right) candidates minus the PDG value as a function of decay vertex
radius. The solid circles are for track reconstruction with the normal
amount of detector material.  The open squares are for track
reconstruction assuming 20\% more material in the SVT.  The dashed lines
show the $\pm1\sigma$ uncertainty on the PDG value for the masses.}
\end{figure*}

The main systematic uncertainty on the \Lc mass comes from
uncertainties in the energy-loss correction in charged particle
tracking. The low \Lz and \KS fitted mass values given in
Table~\ref{tab:crossSignal} indicate that the energy-loss correction
may be underestimated. The long lifetimes of \Lz and \KS hadrons enable
us to study this in more detail. Figure~\ref{fig:lzRadius}
shows the \Lz and \KS fitted mass values as a
function of the radial distance from the interaction point to their
decay point. The further the decay point is from the interaction
point, the less material the charged daughter particles traverse, and
so energy-loss corrections become less significant. The deviation from
the PDG value is seen to be largest for decays closest to the
interaction point, thereby strongly indicating that the
underestimation of the mass values is due to insufficient material
corrections. The lower fitted mass values at radii of 12\,cm, 21\,cm, and
23\,cm coincide with vertices reconstructed inside or very near
material sites. The same effect is observed in the reconstruction of
simulated data. This is related to details of the incorporation of
energy-loss corrections into the track-fit procedure in such
circumstances.

The effect of increasing the amount of material assumed during the
track reconstruction has been studied using several different
scenarios. Figure~\ref{fig:lzRadius} shows what
happens if the material density in all parts of the SVT is increased
uniformly by 20\%. This is a gross simplification; however it removes
most of the dependence on decay radius and moves the fitted hadron
masses closer to their PDG values. The \KS mass is, however,
still consistently low compared to the PDG value by about
0.15\mevcc.

\begin{figure*}
\centerline{\includegraphics{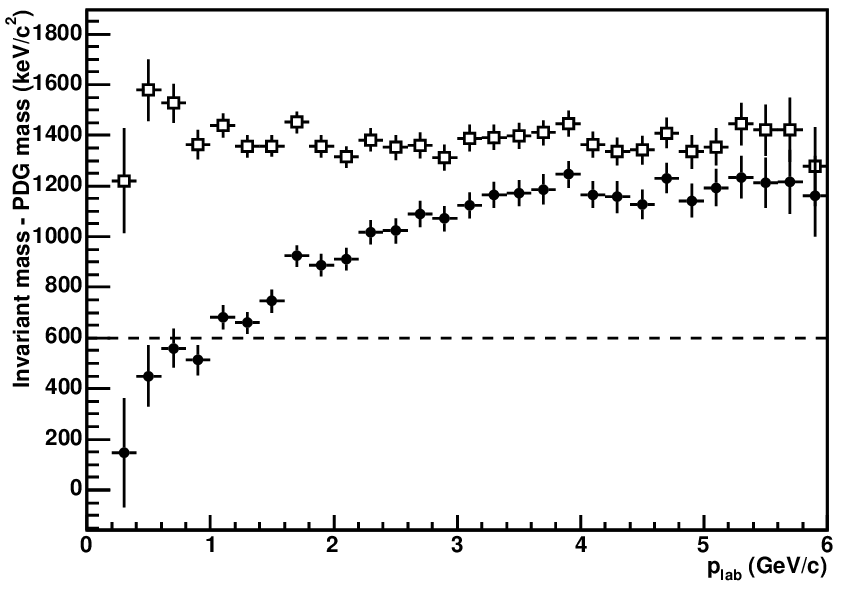} \includegraphics{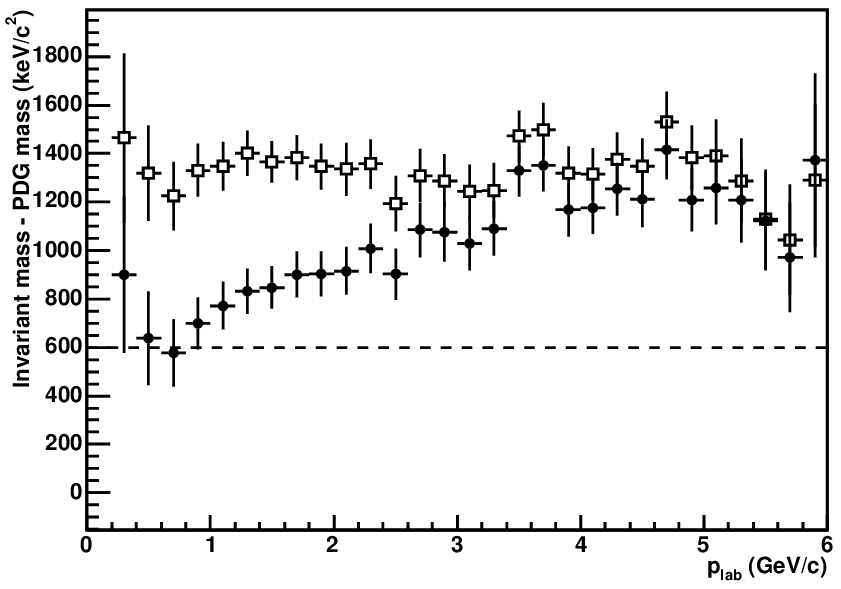}}
\caption{\label{fig:pKpiMomentum} Fitted mass value for \LcpKpi (left) and \LcpKs (right) candidates minus
the PDG value as a function of \Lc laboratory momentum. The solid
circles are for track reconstruction with the normal amount of
detector material.  The open squares are for track reconstruction
assuming 20\% more material in the SVT.  The dashed line shows the
$+1\sigma$ uncertainty on the PDG value of the \Lc mass.}
\end{figure*}

Another way to investigate the energy-loss correction is to study
the fitted mass as a function of particle momentum. The lower the momentum
of a charged particle, the more significant the energy-loss
corrections become. Candidates are reconstructed from multiple tracks,
but to simplify this investigation, the mass value is studied as a function
of net candidate momentum calculated in the laboratory
frame. Figure~\ref{fig:pKpiMomentum} shows
the fitted mass value for \LcpKpi and \LcpKs candidates as a function of the
laboratory momentum.  Above 3\gevc the fitted mass value reaches a limit that is
significantly above the PDG, while below 3\gevc it falls by
more than 800\kevcc as the momentum decreases. Similar behavior is
observed for \Lz and \KS hadrons, but the limiting value is
reached at about 2\gevc. The decrease in mass is only about 60\kevcc for
\Lz baryons and about 500\kevcc for \KS mesons.

Increasing the assumed material density in the SVT by 20\% is seen to greatly reduce
the radial mass dependence in the \Lz and \KS control samples and
the momentum dependence in all control samples. We therefore
apply this change to the reconstruction of the two low
\Qval modes in order to obtain a more accurate mass measurement. 
However, since the \KS mass still is not in agreement with the PDG
value, we will use the largest observed variation in the fitted mass
when we vary the material model as an estimate of the systematic
uncertainty.

\subsection{Magnetic Field Dependence}

\label{sec:magfield}

The momentum measurements, and thus the mass measurements, depend
critically on the magnetic field. The main component of the magnetic
field is the solenoid field, which has an average value of 1.5\,T
parallel to the beam axis. The field was mapped very precisely with
movable Hall probes before the detector was installed; an NMR probe
measured the absolute field strength.
The field strength in the tracking volume is estimated to be known to
an accuracy of about 0.2\,mT. The second most significant field
component comes from the permanent magnets used for the final focusing
and bending of the beams. The magnet closest to the interaction point
is about 20 cm away.  The fringe fields in the tracking volume from
the magnets are weak and have been measured. More uncertain is the
contribution to the magnetization of the permanent magnet material due
to the solenoid field. This effect is measured only at a few specific
points with Hall and NMR probes and we therefore rely on a finite
element calculation to estimate the magnetization effect
elsewhere. This model depends on the permeability of the
magnets. These magnets are made of a SmCo alloy, which has a measured
permeability $\mu=1.07$ in the direction transverse to the solenoidal
field. However this is an average over many samples, which range from
1.04 to 1.10. Furthermore in the direction of the solenoidal field the
permeability of the SmCo elements is measured to be about 5\% larger.
Therefore there is a significant uncertainty on this component of the
field. The longitudinal field component at the interaction point from
these magnets is about $+9$\,mT. This increases the curvature of
charged particle trajectories.

\begin{figure}
\centerline{\includegraphics{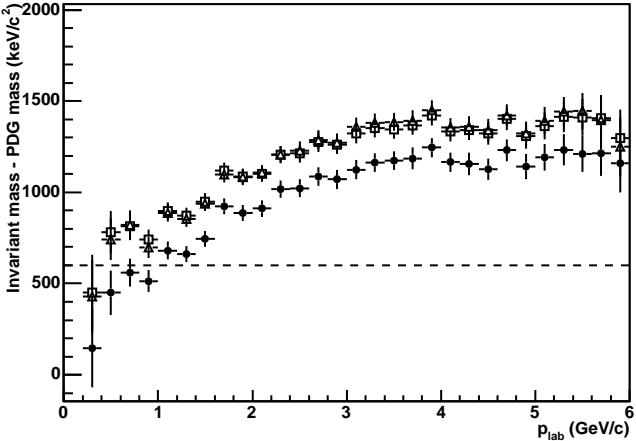}}
\caption{\label{fig:pKpiField} Fitted mass of \LcpKpi candidates minus
the PDG value as a function of \Lc laboratory momentum. 
The additional energy-loss correction has not been applied.
The solid circles are for track reconstruction with the normal
magnetic field.  The open squares are for track reconstruction assuming a
0.02\% higher solenoid field. The open triangles are for track
reconstruction assuming 20\% higher magnetization.  The dashed line shows
the $+1\sigma$ uncertainty on the PDG value of the \Lc
mass.}
\end{figure}

We vary the assumed solenoid field strength by 0.02\% for the systematic
uncertainty studies. The magnetization field is varied by 20\% in
order to account for differences between the direct field measurements
and the permeability measurements.  Fig.~\ref{fig:pKpiField} shows the
effect on the fitted \LcpKpi mass value as a function of momentum. The
fitted mass value is seen to be shifted by the same amount independent
of the momentum.  The same is true for the other control samples, in
particular for the \Lz and \KS mass dependence on decay vertex
radius. The control samples therefore provide little guidance on the
magnetic field uncertainty.

\subsection{$\protect\bm{\mphi}$ Dependence}

\begin{figure*}
\centerline{\includegraphics{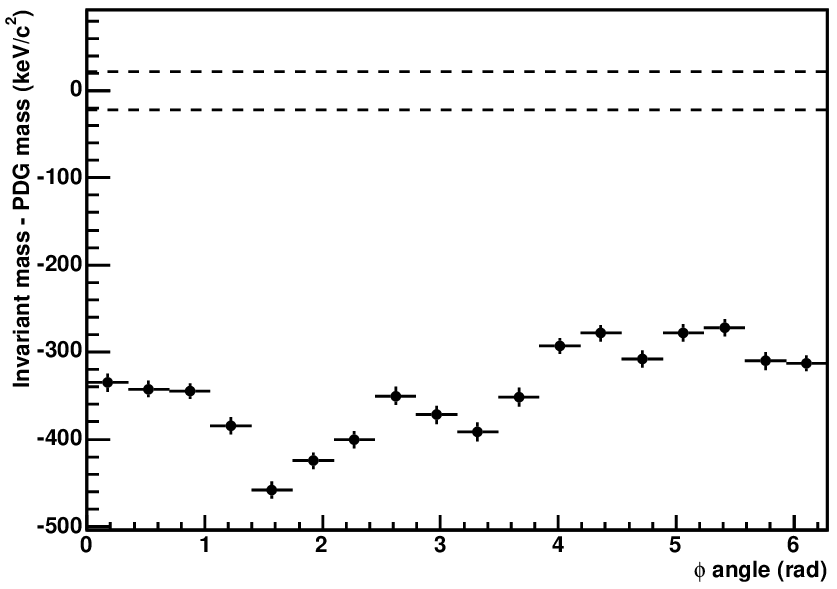} {\includegraphics{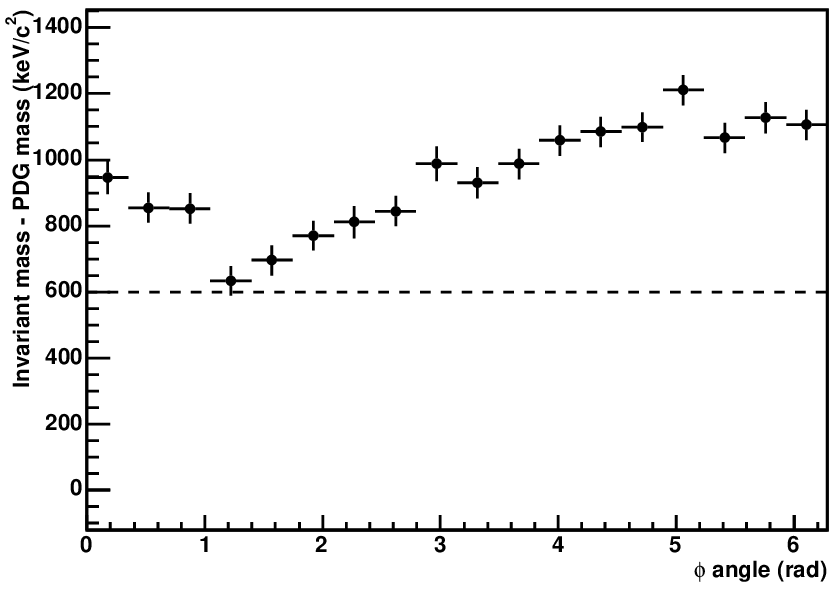}}} 
\caption{\label{fig:ksphi} Fitted mass value of $\KS\to\pipi$ (left) and \LcpKpi (right) candidates minus
the PDG value as a function of the azimuthal direction.
The additional energy-loss correction has not been applied.
The dashed lines show the $\pm1\sigma$ uncertainty on the PDG
value of the masses.}
\end{figure*}

Studies of the control samples reveal a significant dependence of the
fitted mass value on the azimuthal angle \mphi of the hadron candidate
momentum at the origin. This effect is not seen in simulated
events. The \mphi dependence is shown in Figure~\ref{fig:ksphi} for
the \KS and \LcpKpi samples. The effect is seen to be roughly
anti-symmetric in \mphi with the lowest fitted mass at
$\mphi=\frac{\pi}{2}$, corresponding to upward going hadrons, and the
largest fitted mass at $\mphi=\frac{3\pi}{2}$, corresponding to
downward going hadrons.

We have not been able to identify the source of the \mphi dependence,
but we can estimate the potential impact on the \Lc mass
measurement. The magnitude of the \mphi variation increases with the
momentum of the reconstructed hadron. This shows that the variation is
not due to some asymmetry in the material distribution that was
unaccounted for in reconstruction, which would influence low momentum
particles the most.  The \mphi dependence is also observed when only
the DCH is used for reconstructing the tracks. This indicates that the
source may be related to the internal alignment of the DCH, which is
based on end-plate surveys done during construction of the chamber.

The \mphi dependence in the fitted mass can be reproduced
qualitatively in simulated events by introducing an explicit \mphi
dependence in the measured track momenta. To get the largest \mphi
dependence at high momentum, the change has to be introduced as a bias
in the track sagitta measurements. To a good approximation this
corresponds to changing the transverse track momentum $\pt$ according
to
\begin{eqnarray}
\frac{1}{\pt'}&=&\frac{1}{\pt}+\delta\sin \mphi. \label{eq:pdistort2}
\end{eqnarray}
The constant $\delta$ is chosen to reproduce the observed magnitude of the \mphi
dependence.  The chosen value corresponds to a change in track sagitta
of \constDistortSagitta\mum, where the full sagitta for a track with
1\gevc transverse momentum
is about 3\cm. Applying this modification to the \LcLKKs and
\LcSKKs Monte Carlo samples introduces a \mphi dependence in the
fitted mass values with an amplitude of \constLKKsDistortTwo\kevcc and 
\constSKKsDistortTwo\kevcc, respectively. When averaging over all \mphi,
the fitted mass value does not change, but given that the source of the \mphi
dependence is not understood, we use these amplitudes as 
estimates of systematic uncertainty.

%% file: systematics.tex
The major sources of systematic uncertainty (energy-loss correction,
magnetic field, and \mphi dependence) have been described in detail in
the previous section. All of the known systematic uncertainties
are listed in Table~\ref{tab:systematics}. For the
systematic uncertainty on the energy-loss correction we use the
observed change in mass when the material density is increased
uniformly by 10\% in the tracking volume, which is a slightly larger
change than the one from increasing the density in the SVT by 20\%. The
fit-procedure uncertainty summarizes the variation in the fitted mass
value when the shapes of the signal, the background, and
the choice of binning are varied. Possible biases from the internal
alignment of the SVT are studied by applying small distortions to the
SVT alignment in simulated events. The magnitude of the distortions
correspond to the changes observed in the internal alignment between
different run periods. For the \LcSKKs mode, we vary the EMC energy
scale by 5\%, but due to the low energy of the photon and the mass
constraint on the
\Sz, this has little effect on the fitted mass.  Finally the \Lz, \Kp
and \KS masses have uncertainties of 6, 16 and 22\kevcc and the
effect of these uncertainties on the \Lc mass has been estimated.

\begin{table}
\caption{\label{tab:systematics} Systematic uncertainty contributions to the \Lc mass measurements from the \LcLKKs and \LcSKKs samples (in \kevcc).}
\begin{ruledtabular}
\begin{tabular}{lrr}
\multicolumn{2}{r}{\LcLKKs}     & \LcSKKs \\ \hline
Solenoid field          & $\pm \constLKKsPRDSolenoid$ & $\pm \constSKKsPRDSolenoid$ \\ 
Magnetization           & $\pm \constLKKsPRDMagnet$   & $\pm \constSKKsPRDMagnet$ \\ 
Energy-loss correction  & $\pm \constLKKsPRDMaterial$ & $\pm \constSKKsPRDMaterial$ \\
\mphi dependence        & $\pm \constLKKsDistortTwo$  & $\pm \constSKKsDistortTwo$  \\
Fit procedure           & $\pm \constLKKsPRDFitProc$  & $\pm \constSKKsPRDFitProc$  \\
SVT alignment           & $\pm \constLKKsPRDAlign$    & $\pm \constSKKsPRDAlign$    \\
EMC energy scale        &     ---                     & $\pm \constSKKsPRDEScale$   \\
Particle masses         & $\pm \constLKKsPRDMassChg$  & $\pm \constSKKsPRDMassChg$  \\ \hline
Total systematic        & $\pm \constLKKsPRDTotal$    & $\pm \constSKKsPRDTotal$
\end{tabular}
\end{ruledtabular}
\end{table}

\subsection{Energy-Loss Correction}

The fitted \Lc mass values from Section~\ref{sec:signal} need to be
corrected for the underestimated energy loss. The correction is
calculated by increasing the material density of the SVT by 20\%,
which is seen in the control samples to remove most of the momentum
and decay-radius dependence. The corrections are
\constLKKsMassOffset\kevcc and \constSKKsMassOffset\kevcc for the
\LcLKKs and \LcSKKs samples, respectively. This gives the following
results for the \Lc mass:
\begin{eqnarray*}
&&m(\Lc)_{\LKKs} =\\ &&\qquad \constLKKsResult,\\
&&m(\Lc)_{\SKKs} =\\ &&\qquad \constSKKsResult.
\end{eqnarray*}

\subsection{Combined Result}

The systematic uncertainties on the two measurements are highly, but
not fully, correlated. We combine the two mass measurements using the
BLUE (Best Linear Unbiased Estimate) technique~\cite{BLUE}. Besides
the statistical uncertainty, we consider the fit procedure uncertainty
and uncertainties related only to the \LcSKKs mode to be uncorrelated,
while the remaining systematic uncertainties are 100\% correlated.
The correlation coefficient for the two measurements is estimated to
be \constCombCorr. The combined mass result is
\[
m(\Lc)=\constCombMass\mevcc.
\]

\subsection{Mass Cross-checks}

From the two large-\Qval \Lc data samples, we obtain measurements of the \Lc
mass that can be compared to our more precise measurements from the
\LcLKKs and \LcSKKs samples. To keep the syste\-matic uncertainty low,
we use only \LcpKpi and \LcpKs candidates with laboratory momentum
above 3\gevc, as Figure~\ref{fig:pKpiMomentum}
shows that those candidates have less
dependence on the assumed amount of detector material. The resulting
mass value for each decay mode from fitting the invariant-mass spectra
with the sum of two Gaussian distributions with common mean is given in
Table~\ref{tab:pKxSignal}. The central values are corrected for the
shift in mass observed when the material density of the SVT
is increased by 20\%. These corrections are \constpKpiCorrection\kevcc and
\constpKsCorrection\kevcc for the \LcpKpi and \LcpKs modes, respectively.
\begin{table}
\caption{\label{tab:pKxSignal} Fitted \Lc mass values and major contributions to the systematic uncertainty for the \LcpKpi and \LcpKs samples used for cross-checks (in \mevcc).}
\begin{ruledtabular}
\begin{tabular}{lD{x}{\pm}{-1}D{x}{\pm}{-1}}
& \multicolumn{1}{c}{\LcpKpi} & \multicolumn{1}{c}{\LcpKs} \\ \hline
\input PRDtables/PRDpkx.tex
\end{tabular}
\end{ruledtabular}
\end{table}

Table~\ref{tab:pKxSignal} also lists the major systematic
uncertainties. The solenoid and magnetization fields are varied as
for the low-\Qval modes. For the energy-loss correction, we compare the effect
of increasing the density of the SVT by 20\% to the effect of
increasing the density of material in the tracking volume by 10\%,
taking the larger change from the standard reconstruction as an
estimate of the uncertainty. For both decays the larger effect is the
10\% material change in the full tracking volume. The \mphi dependence
is estimated by introducing a \mphi dependence in the simulation as
described by Eq.~(\ref{eq:pdistort2}). The uncertainty is the maximum
change in mass introduced.  The results
\begin{eqnarray*}
&&m(\Lc)_{\pKpi} =\\ &&\qquad \constpKpiResult,\\
&&m(\Lc)_{\pKs} =\\ &&\qquad \constpKsResult.
\end{eqnarray*}
are in good agreement with our main result, but have larger systematic
uncertainties.

The \LcLKKs sample is sufficiently large that we can fit the \Lz and
\KS mass distributions for candidates that are combined to form the \LcLKKs.  In order to 
fit the \Lz and \KS mass distributions, the mass constraint is removed from the
candidate reconstruction and the resulting invariant mass spectra are
fit with a double-Gaussian signal shape and a linear background. The
resulting mass values and systematic uncertainties are listed in
Table~\ref{tab:lzksSignal}. The mass correction for the underestimated
energy loss and the systematic uncertainties are estimated as for the
\LcpKpi and \LcpKs samples. The final results after the energy-loss correction,
\begin{eqnarray*}
m(\Lz)&=&\constMassResultlz\mevcc,\\
m(\KS)&=&\constMassResultks\mevcc,
\end{eqnarray*}
are in agreement with the PDG
\begin{eqnarray*}
m_{\textrm{PDG}}(\Lz)&=&1115.683\pm0.006\mevcc,\\
m_{\textrm{PDG}}(\KS)&=&497.648\pm 0.022\mevcc.
\end{eqnarray*}
Since the \Lz and \KS candidates are the same candidates used in
the final \Lc sample, the agreement with the PDG mass values
gives further confidence in the \Lc mass result.

\begin{table}
\caption{\label{tab:lzksSignal} Fitted \Lz and \KS masses and the major contributions to the systematic uncertainty for candidates from the \LcLKKs sample (in \mevcc).}
\begin{ruledtabular}
\begin{tabular}{lD{x}{\pm}{-1}D{x}{\pm}{-1}}
&  \multicolumn{1}{c}{\Lz} &  \multicolumn{1}{c}{\KS} \\ \hline
Fitted Mass            &  1115.657 x 0.014 &  497.359 x 0.040    \\
Corrected Mass         &  1115.679 x 0.014  & 497.560 x 0.040    \\ \hline
Solenoid field         & x 0.009   & x 0.068   \\
Magnetization          & x 0.005   & x 0.054   \\
Energy-loss correction & x 0.040   & x 0.242   \\
\mphi dependence       & x 0.006   & x 0.059   \\ \hline
Total systematic       & x 0.041   & x 0.264    
\end{tabular}
\end{ruledtabular}
\end{table}

%% file: PRDtables/PRDpkx.tex
Fitted Mass        & 2286.182 x 0.018 & 2286.216 x 0.034 \\
Corrected Mass     & 2286.393 x 0.018 & 2286.361 x 0.034 \\ \hline
Solenoid field     &  x  0.181 &  x  0.196 \\
Magnetization      &  x  0.207 &  x  0.222 \\
Energy-loss correction  &  x  0.278 &  x  0.199 \\
\mphi dependence   &  x  0.217 &  x  0.236 \\ \hline
Total systematics  &  x  0.447 &  x  0.428 \\

%% file: summary.tex
We have presented a precision measurement of the \Lc mass using the
low-\Qval decay modes \LcLKKs and \LcSKKs in order to minimize
systematic uncertainty. The measured mass in the two modes is
\begin{eqnarray*}
&&m(\Lc)_{\LKKs} =\\ &&\qquad \constLKKsResult,\\
&&m(\Lc)_{\SKKs} =\\ &&\qquad \constSKKsResult.
\end{eqnarray*}
Combining these measurements, taking the correlated systematics
into account, the final result for the \Lc mass is 
\[
m(\Lc) = \constCombMass\mevcc.
\]
This result is in agreement with the mass values measured in other \Lc
decay modes, although those are subject to large systematic
uncertainty. The systematic uncertainty has been cross-checked using
large data samples of \Lz, \KS and \Lc decays. The studies have shown
that there is an underestimation of the energy-loss correction, and a
dependence on azimuthal angle in the standard \babar\ track
reconstruction. The impact on the mass measurement has been taken into
account in the corresponding estimates of systematic uncertainty.

This \Lc mass measurement is the most precise measurement of an open
charm hadron mass to date and is an improvement in precision by more
than a factor of four over the current PDG value of
$2284.9\pm 0.6\mevcc$. Our result is about $2.5\sigma$ higher
than the PDG value, which is based on several high \Qval
decay modes, mainly \LcpKpi decays. Theoretical calculations of the
\Lc mass, such as those based upon lattice QCD~\cite{Bowler:1996ws} or advanced
potential models~\cite{Albertus:2003sx}, currently have significantly
larger uncertainty than the experimental result presented here.

%% file: pubboard/acknowledgements.tex
We are grateful for the 
extraordinary contributions of our \pep2\ colleagues in
achieving the excellent luminosity and machine conditions
that have made this work possible.
The success of this project also relies critically on the 
expertise and dedication of the computing organizations that 
support \babar.
The collaborating institutions wish to thank 
SLAC for its support and the kind hospitality extended to them. 
This work is supported by the
US Department of Energy
and National Science Foundation, the
Natural Sciences and Engineering Research Council (Canada),
Institute of High Energy Physics (China), the
Commissariat \`a l'Energie Atomique and
Institut National de Physique Nucl\'eaire et de Physique des Particules
(France), the
Bundesministerium f\"ur Bildung und Forschung and
Deutsche Forschungsgemeinschaft
(Germany), the
Istituto Nazionale di Fisica Nucleare (Italy),
the Foundation for Fundamental Research on Matter (The Netherlands),
the Research Council of Norway, the
Ministry of Science and Technology of the Russian Federation, and the
Particle Physics and Astronomy Research Council (United Kingdom). 
Individuals have received support from 
CONACyT (Mexico),
the A. P. Sloan Foundation, 
the Research Corporation,
and the Alexander von Humboldt Foundation.